# Enhancing Paretic Propulsion Post-Stroke via a Wearable System for Real-Time Unilateral Haptic Feedback of Anterior Ground Reaction Forces

Cameron A. Nurse[1], Kelly Breen[2], Matthew McGuire[2], Sarah Prokup[2], Arun Jayaraman[2], Quentin Sanders[1]

[1] Bioengineering Department, George Mason University, Fairfax VA 22030

[2] Max Näder Center for Rehabilitation Technologies and Outcomes Research, Shirley Ryan AbilityLab, Chicago, IL, USA

## 0. Abstract

Gait rehabilitation interventions targeting paretic propulsion can improve walking speed and function in individual's post-stroke. Previous work has demonstrated that real-time biofeedback targeting anterior ground reaction forces (AGRFs) can increase propulsion in individuals post-stroke, however this work was confined to lab-based treadmills, limiting practical utility. Here we investigate the short-term effects of real-time AGRF gait biofeedback during overground walking using wearable inertial measurement units (IMUs) and a haptic feedback device. Eight individuals with chronic post-stroke hemiparesis completed four 3-minute training bouts. During training, faded haptic biofeedback was provided to increase paretic AGRF during terminal stance. Gait biomechanics were assessed before, during, and after training, and during a retention test conducted without biofeedback after a rest period. The primary dependent variable was peak paretic AGRF, while secondary variables included paretic peak trailing limb angle (TLA), step length and walking speed. Compared to baseline, peak AGRF increased post-feedback and retention tests. Similar trends were observed in TLA, and step length, although these increases were not statistically significant while speed showed a significant change from baseline. Examining individual participants 63% participants (*responders*) increased AGRF at retention, while 37% experienced decreases (*non-responders*). Non-responders had lower physical capability, evidenced by two-minute walk distance at screening and AFO use during training, suggesting this intervention may suit patients with more residual ankle mobility and strength. Nonetheless our results suggest AGRF biofeedback can be implemented in practical settings with wearable systems and is a promising gait training strategy to target propulsive deficits in individuals post stroke.

## 1. Introduction

Stroke is one of the leading causes of chronic disability in older adults, with nearly 800,000 new cases each year and more than 8 million individuals in the United States living with the effects of stroke [1]. A hallmark of post-stroke gait dysfunction is the impaired ability to generate forward propulsion with the paretic limb [2]. Anterior ground reaction force (AGRF), a biomechanical correlate of propulsive force, plays a critical role in advancing the body's center of mass, mediating the transition from stance to swing. In individuals with hemiparesis, reductions in paretic AGRF are associated with slower walking speeds, greater spatiotemporal asymmetry, and more severe motor impairment [3], [4], [5]. Accordingly, restoring paretic propulsion has emerged as a central objective of gait rehabilitation following stroke.

Providing real-time biofeedback of propulsive metrics such as AGRF has demonstrated efficacy in enhancing gait performance in post-stroke populations. Studies have shown that treadmill-based AGRF biofeedback can elicit unilateral increases in paretic propulsion [6], [7]. Beyond AGRF, treadmill-based biofeedback of related gait parameters such as limb loading, step length, and walking speed has also produced meaningful improvements in stroke rehabilitation. For example, Kaźmierczak et al.[8] reported that treadmill training with visual biofeedback, compared to conventional overground training, significantly increased step length, walking speed, and distance covered, while also enhancing static balance and reducing reliance on orthopedic aids after a four-week intervention. Similarly, trailing limb angle (TLA) has been studied as a surrogate marker of propulsion, as it demonstrates moderate correlation with peak AGRF and can be measured with camera-based systems or wearable IMUs [9], [10]. However, biofeedback targeting TLA consistently yields smaller gains in AGRF than direct AGRF feedback [11]. The modest therapeutic effect utilizing biofeedback with surrogate metrics can likely be attributed to compensatory strategies commonly adopted by post-stroke individuals. Since gait recovery often progresses proximally to distally, many individuals rely on preserved hip function to generate propulsion. This hip-dominant strategy is metabolically inefficient compared to balanced contributions from the hip and ankle [12]. TLA based biofeedback primarily targets hip mechanics, however propulsive forces are primarily generated at the ankle. This suggests that while surrogate-based biofeedback can be beneficial, its impact on propulsion will be constrained, but by developing methods to directly target ankle kinetics and propulsion during overground walking we will potentially observe larger improvements.

Recent advances in wearable sensing and machine learning offer promising avenues to address this need. IMUs are lightweight, inexpensive, and unobtrusive, making them well-suited for capturing kinematics in naturalistic environments outside of specialized laboratories. Prior studies have shown that IMUs can be used to estimate GRFs but many approaches have relied on regression models or physics-based inverse dynamics, which are sensitive to modeling assumptions and often lack generalizability across populations [13]. Others have required subject-specific calibration or force plates for model training, limiting their clinical utility [14]. Machine learning techniques, particularly deep learning, can overcome many of these limitations by capturing nonlinear relationships between limb kinematics and GRFs. Such models are capable of learning from heterogeneous data without subject-specific calibration and can generalize across varied gait patterns and levels of impairment. Our group has developed a deep learning model that estimates AGRF in real time using IMUs during overground walking in individuals post-stroke [15], enabling the delivery of targeted AGRF biofeedback in clinically relevant and scalable settings.

By integrating wearable IMU-based sensing with machine learning driven AGRF estimation, we sought to address the limitations of laboratory-bound treadmill systems and enable scalable, clinically accessible gait rehabilitation in real-world environments. In the present study, we evaluate the efficacy of haptic biofeedback targeting paretic AGRF during overground walking in individuals' post-stroke. We hypothesize that AGRF-directed feedback will enhance paretic propulsion, kinematic propulsion surrogates (TLA and step length) and walking speed relative to a no-feedback control condition. This work seeks to advance gait rehabilitation by providing a clinically feasible, and translatable biofeedback paradigm that extends beyond the confines of treadmill-based interventions

## 2. Methods

### 2.1 Wearable biofeedback device and methodology

We designed a wearable system that enables real-time, overground biofeedback without the need for treadmill or laboratory-grade equipment, facilitating more naturalistic gait training. AGRF biofeedback was delivered via a custom haptic armband worn on the unaffected upper arm and consisted of two vibration motors controlled by a Raspberry Pi Pico microcontroller (Fig. 1A). The device was lightweight and unobtrusive, allowing unrestricted overground walking. A custom MATLAB script streamed IMU data from the Movella Awinda system (Movella, Enschede, Netherlands), calculated AGRF using the machine learning model described in Section 2.2 and sent control signals to the armband via WiFi User Datagram Protocol (UDP) in real time. The system operated at a data rate of 50 Hz, with low latency to ensure immediate feedback.

Haptic feedback was delivered as a double pulse vibration whenever the estimated AGRF exceeded a participant-specific threshold, set at 5% above the average peak AGRF from their baseline trial; this value was informed by able-bodied and post-stroke pilot testing, which indicated that smaller targets maximized participants' ability to achieve consistent success. During the biofeedback conditions, the participant's AGRF was estimated continuously, if the AGRF signal was at least 5% greater than the baseline a success flag was triggered and vibrational feedback was provided at the initiation of swing phase (when AGRF dropped to zero), indicating success (Fig 1B). Feedback was provided using a faded dosing schedule [30], with the 3-min biofeedback bouts interspersed with increasing no-feedback walking periods.

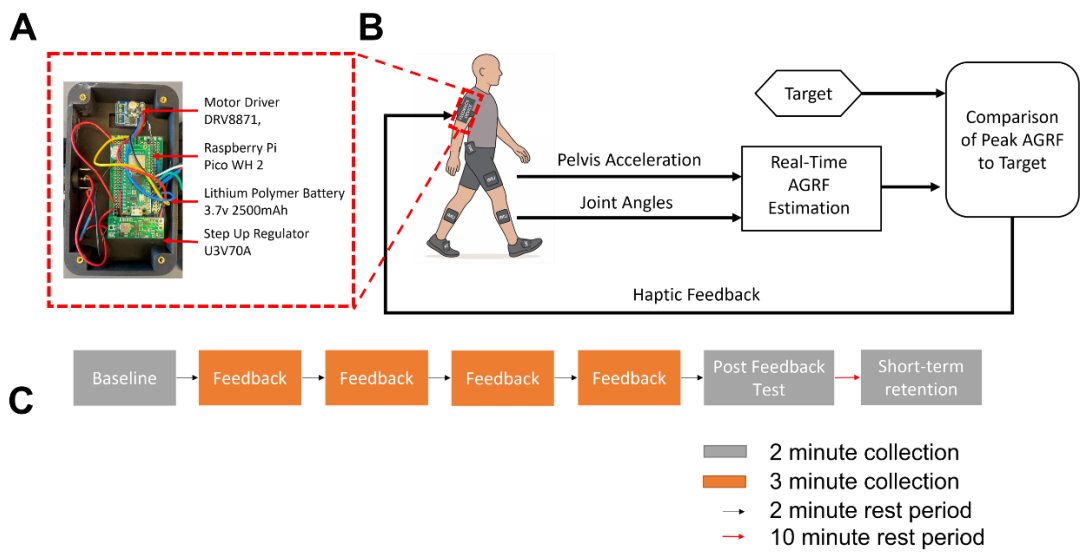

**Fig 1**: Haptic biofeedback was given based on a real-time estimate of anterior ground reaction force. The device is depicted in **A.** The signal pathway is overviewed in **B.** The intervention protocol is described in **C.**

2.2 Machine learning model development

We used a hybrid convolutional-recurrent neural network (CNN-LSTM) trained on stroke walking data from a prior publication [15]. The model takes time-series IMU data from sensors positioned on the pelvis, thighs, shanks, and feet as input, and continuously estimates AGRF over time. This hybrid architecture was chosen to capture both short-term spatial patterns and longer-term dependencies in sequential gait data.

The network begins with a 1D convolutional layer containing 128 filters with a kernel size of 5 and Rectified Linear Unit (ReLU) activation, which extracts local temporal features from the input sequences. The convolutional output is then passed to a single LSTM layer with 64 hidden units to capture longer-range temporal dependencies. To reduce overfitting, a dropout layer with a rate of 0.3 follows the LSTM. Finally, two fully connected (dense) layers are applied: the first with 32 units and ReLU activation, and the second with a single output neuron to produce the continuous AGRF prediction.

The model uses a sliding window of the current and five previous time frames of IMU data, ensuring low computational latency and compatibility with embedded systems. This design avoids reliance on future data, enabling true online, frame-by-frame inference suitable for real-time biofeedback. Using a full lower body IMU configuration (pelvis, thighs, shanks and feet), the model achieved a mean absolute error of 0.033 body weights and a correlation of 0.81 for peak AGRF values compared to gold-standard force plate measurements. For full details on the model, refer to the prior publication [15].

2.3 Biofeedback intervention testing

Eight individuals in the chronic phase of stroke (6 males, 2 females; mean age = 56.4 ± 8.7 years, Table 1) were recruited from the Shirley Ryan Ability Lab. Participants wore the Movella Awinda system in the lower-body configuration (pelvis, thighs, shanks, feet). All provided written informed consent prior to participation. The study was approved by the institutional human participants ethics committee of both George Mason University and Northwestern University. Eligible individuals were adults aged 30–80 years with hemiparesis who were at least six months post-stroke. Participants were required to be ambulatory, with or without the use of a cane or walker, and able to walk continuously for six minutes at a self-selected speed without orthotic support and without experiencing significant pain ($\leq 5$ on the visual analog pain scale). Additional criteria included the absence of pain or discomfort in the legs or chest during walking, intact communication and attention abilities (NIH Stroke Scale score $\leq 1$ for language and attention), cognitive function within normal limits (Montreal Cognitive Assessment score > 22), and willingness and ability to comply with all study procedures.

**Table 1: Participant Demographics**

| Age | Sex | Height (cm) | Weight (kg) | Hemiparetic side | Time post stroke (Days) | 2 Minute Walk Test (m) |
|---|---|---|---|---|---|---|
| 68 | M | 179 | 92.7 | R | 1347 | 50.8 |
| 65 | M | 174 | 73.8 | R | 1866 | 115.6 |
| 49 | F | 170 | 106.8 | L | 1956 | 55.2 |
| 65 | F | 160 | 81.5 | L | 5670 | 15.4 |
| 43 | M | 169 | 71.5 | L | 1439 | 72.9 |
| 43 | M | 180 | 73.6 | L | 800 | 74.3 |
| 57 | M | 174 | 80.8 | L | 1196 | 90.2 |
| 53 | M | 176 | 83 | R | 281 | 63.1 |

Participants completed seven overground walking trials at the Shirley Ryan Ability Lab. Control trials lasted 2 minutes, while feedback trials lasted 3 minutes, with 2 minutes of seated rest between each trial. A licensed physical therapist followed each participant with a gait belt for safety, and an experimenter recorded distance walked at 1-min intervals.

All participants began with a baseline 2-min control trial. They were then fitted with a haptic biofeedback armband on the upper arm of their unaffected side. Before beginning the feedback trials, participants received scripted instructions that explained the biofeedback interface, emphasized that the goal was to "push harder" with the paretic leg, and clarified that feedback targeted the paretic side. To supplement these instructions, participants viewed a visual representation of AGRF [11] to illustrate "pushing harder" during forward stepping. Each 3-min feedback trial was divided into alternating feedback and no-feedback periods. During feedback periods, participants received haptic vibration when successful; during no feedback periods, no vibration was delivered even if they were successful. We used a faded feedback approach [16], [17], in which the proportion of time with active feedback decreased across minutes, from 45 sec. feedback and 15 sec. no-feedback in the first minute, to 15 sec. feedback and 45 sec. no-feedback in the final minute.

After completing the feedback trials, the device was removed, and participants performed a second 2-min control trial. Following a 10-min seated rest, participants completed a third 2-min control trial to assess short-term retention effects of the biofeedback training (Fig. 1C). The primary dependent variable was peak AGRF of the paretic leg. Secondary variables included peak TLA, step length, and distance covered during a 2-minute walking test. Step length and TLA were computed from pelvis–foot positions using established methods [9]. Gait events were identified based on the anterior–posterior position of the paretic foot relative to the pelvis, where peaks in this signal corresponded to foot contact and valleys indicated the initiation of the swing phase [18]. Individual stance events of the paretic leg were segmented and time-normalized to 0–100.

Analysis metrics including peak AGRF, step length, and TLA at the time of peak AGRF were extracted for each stance phase. The metrics were averaged across each stance phase to obtain a single representative value per trial. Additionally, gait speed was estimated by using the total distance covered during the first two minutes of walking to assess functional walking capacity. All primary and secondary parameters were analyzed using a repeated measures ANOVA to examine differences across walking conditions (Baseline, Feedback, 2 minutes post-feedback, Short-term retention) using RStudio (PBC, Boston MA, USA). We then assess the effect size (Cohen's d) and absolute difference confidence interval comparing baseline to feedback, 2 minutes post-feedback and short-term retention trials.

Exploratory analyses were also conducted in MATLAB (Mathworks, Natick MA, USA) to assess potential factors influencing responsiveness to the training. These included characteristics of the training feedback, such as the time to first feedback trigger, the total number of feedback triggers per session, the maximum number of consecutive feedback triggers and the coefficient of variation of consecutive triggers during feedback active times as well as participant characteristics, including demographic and functional measures (e.g., age, baseline gait metric, etc.), to explore potential relationships with positive or negative responses to the intervention.

## 3. Results

Group-level changes were observed across all outcome measures during and after the feedback period (Table 2). For AGRF, average increases were observed during feedback (+12.5%), post feedback (+7.4%), and at short-term retention (+15.4%) with small effect sizes observed (Cohen's d ≤ 0.16). For TLA, participants showed increases during feedback (+15.3%), post feedback (+12.6%), and short-term retention (+18.3%). Effect sizes ranged from small to moderate (d = 0.22–0.31). Step length increased by 5.8% during feedback, 4.7% post feedback, and 11.2% at short-term retention. Effect sizes were moderate, with the largest observed at retention (d = 0.49). For walking speed, participants demonstrated increases during feedback (+7.6%), post feedback (+15.3%), and short-term retention (+16.1%). Effect sizes were small to moderate (d = 0.16–0.33). While these trends suggest potential benefits of the intervention, the small to moderate effect sizes and confidence intervals that include zero indicate that these findings should be interpreted with caution, as they may reflect limited reliability. Further the repeated measures ANOVA for paretic peak AGRF showed no significance when comparing pre and post feedback trials (p = 0.59, Table 3). Additionally, the ANOVA for paretic TLA (p = 0.18 Table 3), and step length (p = 0.13, Table 3) was not statistically significant. Gait speed, however, did show significance across trials (p < 0.01, Table 3).

The lack of significance and large effect size could be due to the small sample size coupled with the variability in patient response to the intervention. While the group-level analyses of peak AGRF were not significant, examination of individual data revealed variability in response to feedback. Five participants demonstrated increases in paretic peak AGRF during short term retention (+ 30.21%) compared to baseline (Fig. 3), whereas three participants showed slight decreases (-9.30%) compared to baseline (Fig. 3). A similar trend was observed for TLA (Fig. 3), step length (Fig. 3), and speed (Fig. 3). In some cases, participants showed increases in these measures without corresponding increases in peak AGRF, indicating potential use of compensatory strategies like hip hiking to increase leg swing distance. Percent change in paretic peak AGRF from baseline to the short-term retention was moderately correlated to success metrics during feedback training. The percent change was negatively correlated with time to first feedback trigger (r = –0.50, Fig 4) and the variance of successive triggers (r=-0.40, Fig.4), while positively correlated with both the maximum number of successive triggers (r = 0.59, Fig 4) and the average number of triggers per trial (r = 0.62, Fig 4). No meaningful differences were observed between responders and non-responders for age, height, weight or sex. However, some functional differences were noted, all non-responders used an ankle-foot orthosis (AFO) during data collection, whereas only one responder used an AFO. Additionally, during screening, responders demonstrated a greater 2-minute walk distance (108.5 m) compared to non-responders (60.5 m). These findings suggest that responsiveness to training, and subsequent retention may depend not only on in-session performance but also on baseline functional capacity and reliance on orthoses and walking aids.

**Table 2**: Percent change, effect size and absolute confidence interval compared to baseline show small to moderate impact of biofeedback training on gait parameters

| AGRF | | | | TLA | | | |
|---|---|---|---|---|---|---|---|
| Condition | Avg. % Change | Cohen's d | CI | Condition | Avg. % Change | Cohen's d | CI |
| During Feedback | +12.50 | 0.16 | [-0.04, 0.03] | During Feedback | +15.27 | 0.31 | [-6.02, 3.31] |
| Post-Feedback | +7.41 | 0.06 | [-0.04, 0.04] | Post-Feedback | +12.64 | 0.22 | [-5.42, 3.57] |
| Short-Term Retention | +15.39 | 0.06 | [-0.04, 0.03] | Short-Term Retention | +18.32 | 0.22 | [-7.40, 3.97] |

| Step Length | | | | Speed | | | |
|---|---|---|---|---|---|---|---|
| Condition | Avg. % Change | Cohen's d | CI | Condition | Avg. % Change | Cohen's d | CI |
| During Feedback | +5.81 | 0.34 | [-0.07, 0.04] | During Feedback | +7.55 | 0.16 | [-0.38, 0.28] |
| Post-Feedback | +4.69 | 0.27 | [-0.08, 0.05] | Post-Feedback | +15.25 | 0.32 | [-0.45, 0.24] |
| Short-Term Retention | +11.18 | 0.49 | [-0.1, 0.04] | Short-Term Retention | +16.05 | 0.33 | [-0.46, 0.24] |

**Table 3:** Average metric of peak AGRF, TLA, step length and speed show group level increases during and post feedback

| Metric | Baseline | During feedback | Post feedback | Short-term retention | p-value |
|---|---|---|---|---|---|
| **Peak AGRF (BW)** | 0.088 ± 0.037 | 0.092 ± 0.031 | 0.090 ± 0.028 | 0.095 ± 0.027 | 0.59 |
| **TLA (deg)** | 13.04 ± 3.86 | 14.40 ± 5.03 | 13.97 ± 4.95 | 16.54 ± 3.38 | 0.18 |
| **Step length (m)** | 0.256 ± 0.040 | 0.271 ± 0.055 | 0.271 ± 0.070 | 0.287 ± 0.079 | 0.13 |
| **Speed (m/s)** | 0.64 ± 0.29 | 0.70 ± 0.32 | 0.74 ± 0.34 | 0.75 ± 0.33 | <0.01 |

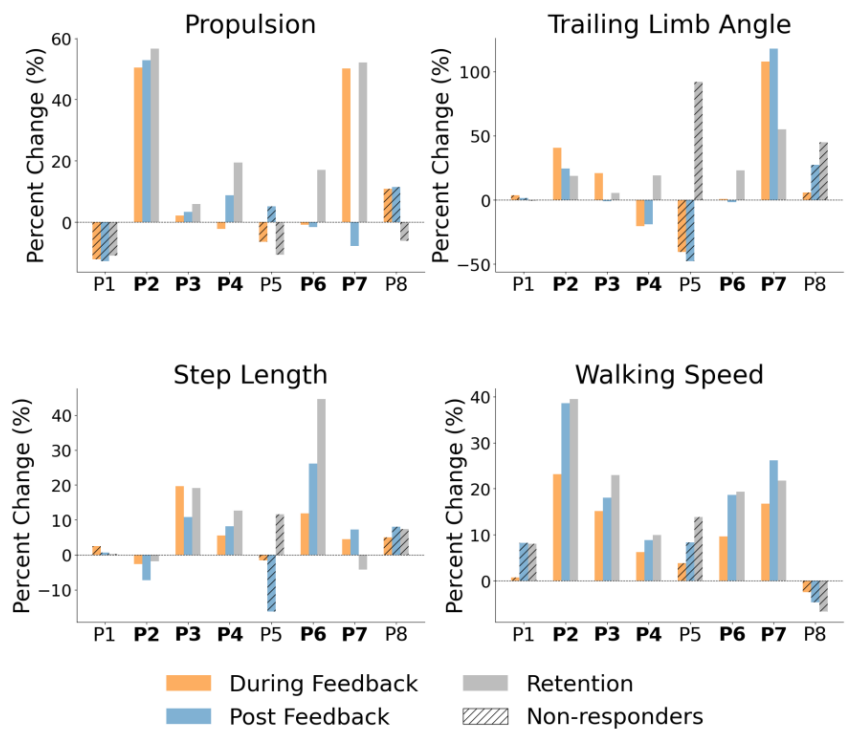

**Fig. 2:** Percent change from baseline in AGRF, TLA, step length, and speed for each participant (P1–P8) across trials.

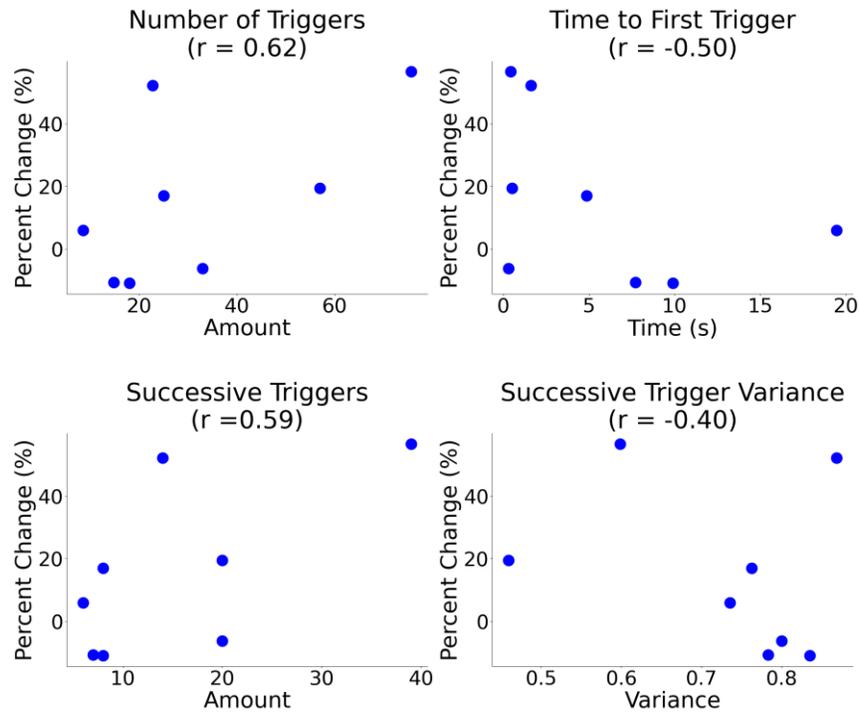

**Fig 3**: Participants' ability to achieve success triggers quickly and consecutively was moderately correlated to increases in propulsion during short term retention compared to baseline

## 4. Discussion

The goal of this study was to evaluate whether haptic biofeedback targeting peak AGRF during walking could effectively modulate propulsion in individuals' post-stroke. We hypothesized that participants would increase peak AGRF and related measures such as TLA, step length, and walking speed compared to baseline walking when provided with direct biofeedback on peak AGRF. Most participants demonstrated increases in AGRF and kinematic surrogates both during and after feedback, partially supporting the hypothesis that direct biofeedback can enhance propulsion, although these changes were not significant at the group level. As with many clinical interventions, several nuances and contextual factors influenced the outcomes. Below, we discuss the major findings of this study, along with key considerations, challenges, and opportunities for future advancement.

We implemented a haptic biofeedback intervention to increase peak AGRF and observed moderate to large increases in kinematic and spatiotemporal metrics, with smaller but variable improvements in AGRF at the group level. Interestingly, increases in kinematic measures such as TLA, step length, and walking speed did not always correspond to increases in AGRF. This observation aligns with previous work comparing TLA- versus AGRF-focused biofeedback in post-stroke gait, which demonstrated that changes in kinematics do not necessarily translate to increases in ankle-centered kinetics or propulsion [11]. Together, these findings suggest that improvements in gait kinematics alone may be insufficient to drive meaningful increases in propulsive force. It is possible that some participants emphasized kinematic strategies such as longer steps or faster walking speed that are correlated with propulsion but do not produce a one-to-one increase in AGRF, resulting in substantial kinematic changes but minimal gains in actual propulsive force.

Variability in AGRF improvements may also reflect methodological differences between our study and prior work. Previous studies using visual, or audiovisual biofeedback during treadmill walking reported more consistent increases in propulsion across participants [7], [11], [19], [20]. Our haptic feedback delivered binary success cues, which provided less graded information than continuous visual displays and required participants to explore strategies to achieve success. The findings of the secondary analysis support this, as participants who achieved success feedback more quickly and consistently showed greater improvements in post-feedback and short-term retention trials. While this may reflect individual capacity to generate increased propulsion, it could also indicate that using only haptic feedback lacked detailed information to guide precise adjustments. Nonetheless, haptic feedback remains a practical option for real-world implementation, as tactile cues can be delivered via small wearable devices or smartphones without requiring the user to focus visually, unlike augmented reality or head-mounted displays which could become distracting. Walking environment may have also played a role: treadmill walking enforces fixed pace and consistency, whereas our overground design introduced natural variability in speed and stride. Finally, the measurement method could contribute to variability. Previous studies measured AGRF directly via force plates, whereas we used a wearable sensor with a machine learning algorithm to estimate AGRF. Although the model performed well during validation [15] estimation errors or under-sampled gait patterns could have produced variable or sporadic feedback, contributing to the inconsistent group-level improvements. Overall, these findings highlight both the challenges and potential of implementing haptic biofeedback in real-world settings. While methodological factors could have contributed to smaller and more variable AGRF improvements compared to previous work, our study demonstrates the feasibility of a wearable, out-of-the-lab system capable of increasing propulsive forces during post stroke gait.

In addition to methodological differences, patient characteristics may also contribute to differences in outcomes. Clinical biofeedback studies, including our own, typically involve small sample sizes (8–12 participants), making direct comparisons across studies challenging. Within our study, participants demonstrated varied responses to AGRF biofeedback, with three of eight participants showing decreases in peak AGRF at short-term retention relative to baseline, classified as non-responders. Notably, the average distance covered during the 2-minute walk test at screening was substantially lower for non-responders (60 m) compared to responders (108 m), suggesting that baseline physical capability may influence retention. Additionally, all non-responders wore an AFO during the experimental protocol. Plantarflexion in late stance is crucial for generating propulsive force during gait, with both TLA and ankle plantarflexion moment contributing significantly to propulsion [10], [21]. Limiting plantarflexion with AFOs could therefore reduce the ability to generate optimal propulsive forces. While AFO use and baseline ability appear influential, other factors such as neuroplasticity, motivation, and intervention design may also contribute to variability [21].

Given that AFOs may restrict propulsive force generation, future AGRF biofeedback interventions might consider reducing or removing AFO use. This could allow greater plantarflexion and larger benefit from propulsive based feedback but would increase fall risk for lower-capability patients. To mitigate these risks, future work could explore coupling propulsive biofeedback with assistive technologies, such as exoskeletons [22] or functional electrical stimulation (FES), to either support safe training without AFOs or augment propulsion directly. Coupling biofeedback with assistive devices, may enhance gait rehabilitation by promoting more natural and effective movement patterns. Biofeedback provides real-time information on gait parameters, allowing individuals to adjust their movements, while assistive devices can supply powered support to the lower limbs, facilitating correct movement patterns and improving engagement [23]. Similarly, FES can stimulate muscle contractions in a targeted manner, and when combined with biofeedback, stimulation can be adapted in real time to optimize propulsion and gait symmetry [24]. Adaptive control systems that integrate voluntary movement, robotic assistance, and FES can further encourage active participation while reducing fatigue [25]. These combined approaches may help overcome limitations of lower capability patients, allowing users to practice propulsion without compromising safety. Building on these propulsion-specific interventions, wearable sensor-based biofeedback devices offer an additional pathway to enhance clinical gait training.

Propulsion-targeted biofeedback has broad clinical and translational potential. In rehabilitation settings, clinicians could use this approach to reinforce optimal push-off strategies, tailor targets to baseline capability, and track performance continuously. This can help therapists deliver highly individualized gait training, reinforcing optimal push-off strategies and encouraging active engagement in propulsion-focused exercises. Beyond the clinic, wearable propulsion biofeedback devices could support home- and community-based gait training. Patients can engage in high-repetition of propulsion-specific walking guided by clinician-prescribed targets, enhancing motor recovery through task-specific, goal-directed practice [26]. Home-based use also enables continuous collection of propulsion metrics, creating rich datasets for longitudinal monitoring and adaptive therapy. The ability to provide skilled, clinician-selected cues without the constant presence of a therapist also expands tele-rehabilitation opportunities. Telehealth protocols have been shown to be non-inferior to in-person therapy for the upper extremity post stroke [27], but comparable approaches for lower extremities are limited. Propulsion-specific biofeedback could fill this gap by supporting both remote gait retraining and performance evaluation, as data collected through home-based devices can be transmitted directly to clinicians for ongoing monitoring and adjustment of therapy plans.

This study had several limitations that should be considered when interpreting the findings. First, the short-term study design restricted our ability to assess long-term retention of learned propulsion strategies and whether gains in AGRF would translate into meaningful clinical outcomes such as walking endurance, community ambulation, or reduced fall risk. Future studies should incorporate repeated practice and longitudinal follow-up to evaluate persistence of motor adaptations and functional carryover. Second, the relatively small sample size may have limited statistical power and the ability to detect subgroup-specific effects (e.g., responders vs. non-responders). Larger cohorts would enable stratified analyses to determine whether factors such as baseline gait impairment, physical capacity, or walking aid use predict responsiveness to propulsion biofeedback. Third, participants received a brief exposure to the intervention (12 minutes of training), prior work demonstrated that short exposures can elicit significant improvements in propulsion [7] during single session testing. Nonetheless, while this was sufficient to examine feasibility and short-term responsiveness, it may underestimate the potential benefits of a sustained training paradigm. Future studies should therefore evaluate the effects of extended and repeated training sessions, ideally integrated into standard rehabilitation programs, to determine the cumulative impact of AGRF biofeedback on propulsion mechanics and overall walking function.

## 5. Conclusion

This study demonstrated the feasibility of a wearable, haptic biofeedback system designed to target AGRF and thereby modulate propulsion in individuals' post-stroke. While group-level improvements in AGRF were modest and variable, most participants increased propulsion, kinematic and spatiotemporal measures during and after training, highlighting both the potential and the challenges of propulsion-specific biofeedback. The unique aspects of this work, direct targeting of propulsive force, delivery through haptic-only binary feedback, and use of machine learning to estimate AGRF from wearable kinematic sensors—underscore its novelty and translational potential compared to previous lab-based approaches. Importantly, variability in participant responses suggests that baseline physical capacity, orthotic use, and feedback design could influence outcomes, pointing to critical directions for future refinement. Overall, this work provides an initial step toward real-world, out-of-the-lab propulsion biofeedback interventions that could complement existing gait rehabilitation strategies and ultimately support greater independence for individuals post-stroke.